\documentclass[slac_one]{revtex4}
\usepackage{graphicx}
\usepackage{fancyhdr}
\newcommand{\MET}{$\not\!\!E_T$}

\begin{document}

\title{{\small{Hadron Collider Physics Symposium (HCP2008),
Galena, Illinois, USA}}\\ 
\vspace{12pt}
Early Standard Model Measurement and Determination of Standard Model Background for Searches} 

%

\author{S. Beauceron\\
on behalf of ATLAS and CMS collaboration}
\affiliation{CERN, CH-1211 Gen\`eve 23, Switzerland}

\begin{abstract}
The Large Hadron Collider (LHC) at CERN in Geneva (Switzerland) will
go in operation in the coming months and will soon enable us to
analyze the highest energy collisions ever produced at an
accelerator. Beyond Standard Model searches at LHC require a detailed
understanding of the detector performance, reconstruction algorithms
and triggering. Precision measurements of Standard Model processes are
also mandatory to acquire the necessary knowledge of Standard Model
background. Both ATLAS and CMS efforts are hence addressed to
determine the best calibration candles and to design a realistic plan
for the initial period of data taking.
\end{abstract}

\maketitle

\thispagestyle{fancy}


\section{INTRODUCTION} 
ATLAS~\cite{Atlas} and CMS~\cite{CMS} detector are two general
detectors which have been designed in order to scrutinize
proton-proton collisions of LHC. The major goal of these experiment is
to search for beyond Standard Model processes and/or to push limits of
Standard Model theory. The first steps of these searches will be to
establish with precision Standard Model processes for a center of mass
of 14~TeV. Within the statistics delivered by LHC over the first years
of running, precised Standard Model measurements will constraint
beyond Standard Model theories and allow us to understand these
background for searches. In this paper, the Standard Model
measurement will be presented as a function of increase luminosity. At
each stage, these measurements will be exposed as a background
for a given search.

Before collisions, the commissioning of the detectors is crucial to
already understand their response. Then with 10pb$^{-1}$ of recorded
collisions, detector synchronization, alignment of detectors and
commissioning of first physics objects will be done. The first physics
will be then dominated by jet physics. With less than 100pb$^{-1}$,
measurement of Standard Model processes using leptons can be addressed
with a high precision and allow the start of the searches. Studies of
complex final states such as $t\bar{t}$ production will help the
finalization of the commissioning period. Beyond 1000pb$^{-1}$, the
area of searches will begin.

\section{FIRST PHYSICS USING JETS}

\begin{figure*}[t]
\centering
\includegraphics[width=135mm]{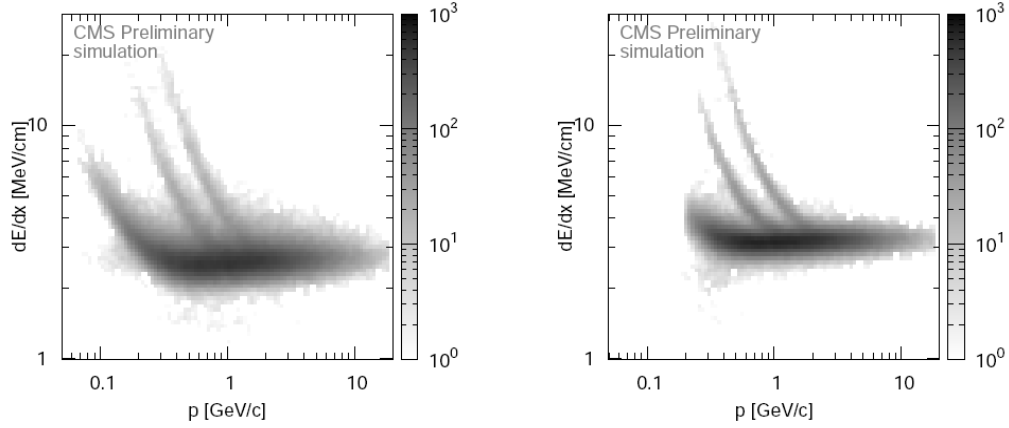}
\caption{Distribution of the truncated mean estimator d$E$/d$x$ as a 
function of momentum $p$ for the pixels hits (left) and strip hits 
(right) of the CMS detector.} 
\label{fig:HadSpec}
\end{figure*}

The first physics events that will be recorded by the detectors will
be mainly minimum bias events. The minimum bias events will be used at
the first stage to calibrate and align detectors. In the meantime a
first look at the charged hadron spectrum at center of energy of
14~TeV will be possible. In the Fig.~\ref{fig:HadSpec}, ones can see
the different $\frac{dE}{dX}$ from proton, kaons and pions for two of
the CMS tracker detector.

\begin{figure*}[t]
\centering
\includegraphics[width=135mm]{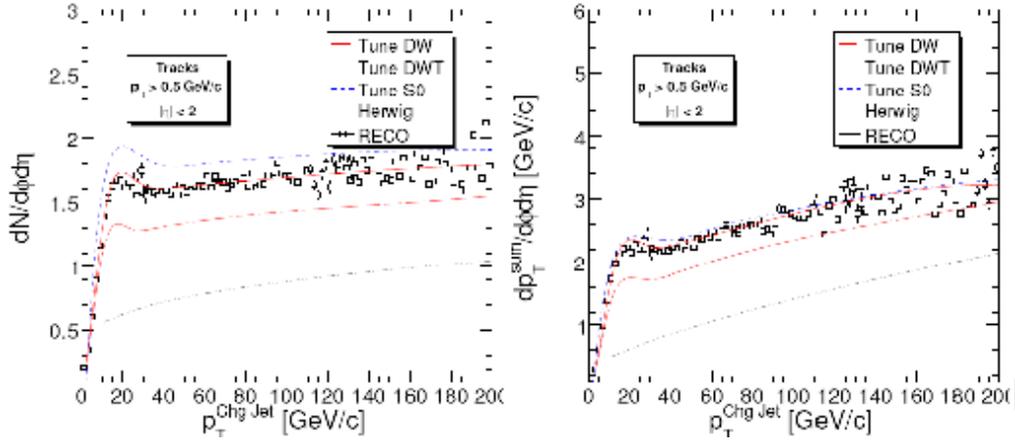}
\caption{Densities dN/d$\eta$d$\phi$ (left) and d$p_T ^{sum}$/d$\eta$d$\phi$ 
(right) for tracks with $p_T > 0.5~GeV/c$, as a function of the leading 
charged jet $p_T$, in the transverse region, for an integrated luminosity 
of 100pb$^{-1}$ collected by CMS detector.}
\label{fig:UE}
\end{figure*}

In parallel, the studies of underlying events are mandatory at start
up. The current simulation of these events is based on an
extrapolation of the Tevatron energy at 2 TeV up to LHC energy at
14~TeV. These extrapolations lead to different values as shown in
Fig.~\ref{fig:UE}. It is them important to be able to discriminate
between different extrapolations in order to fine tune the simulation
as such events will play an crucial role in isolation criteria to
defined isolated leptons.

Jets will be the first reconstructed objects at startup but these are
also the most difficult objects to fully understand. It has to give a
good description of jet/parton properties from an interpretation of
calorimeters response. The response of the calorimeters will be
influenced by experimental factors and physics factors. The main
experimental factors that ones has to take care when building a jet
are the amount of dead material in front of the calorimeters,
longitudinal leakage and lateral shower size, non linearities in the
read out as well as the non-compensated behavior of the
calorimeter. The main physics factors are the understanding of
initial/final state radiation, fragmentation, the amount of underlying
events/minimum bias events.

\begin{figure*}[t]
\centering
\includegraphics[width=135mm]{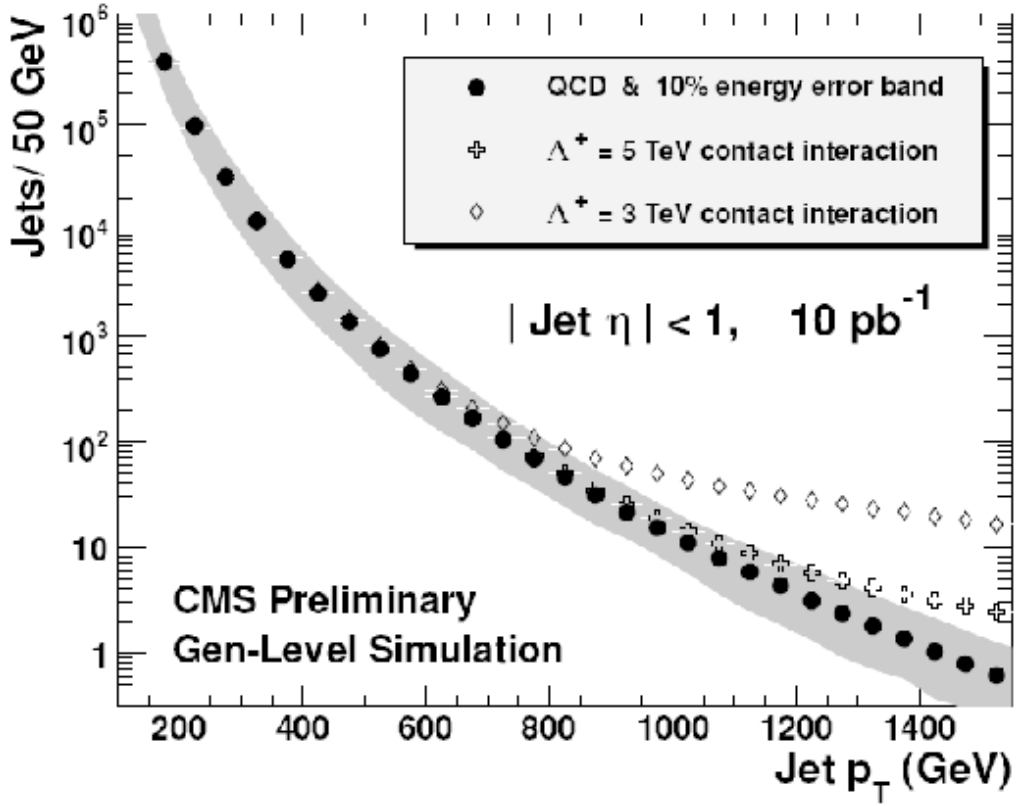}
\caption{Rate of Jets within the central part of CMS detector as a 
function of energy for 10~pb$^{-1}$ of expected integrated luminosity. 
Contact interaction will increase the rate at large energy.} 
\label{fig:ContactInteraction}
\end{figure*}

Nevertheless, the physics program with jets is large and already some
searches/complementary measurement of the Standard Model can be
established. One of the first measurement can be the differential
cross section of jets production which will allow a test of QCD
theories: high momentum objects can be influenced by processes beyond
Standard Model. The contact interaction will tend to enlarge the rate
of production of jets at higher momentum as shown in
Fig.~\ref{fig:ContactInteraction}. CMS
analysis~\cite{CMSContactInteraction} describes with 10 pb$^{-1}$ of
data collected a sensitivity to $\Lambda^+ >$ 2.7 TeV which is the
current limit of Tevatron experiment.

The first advance event topology studied will be the studies of dijet
events and mainly the invariant mass of dijet events. The dijet
response will be used at start up in order to provide a first jet
energy scale. This scale is mandatory to perform studies of invariant
mass but, already, with a few of inverse picobarns of integrated
luminosity, some tests beyond Standard Model can be performed.

\section{CHALLENGING MISSING E$_{T}$}

\begin{figure*}[t]
\centering
\includegraphics[width=72mm]{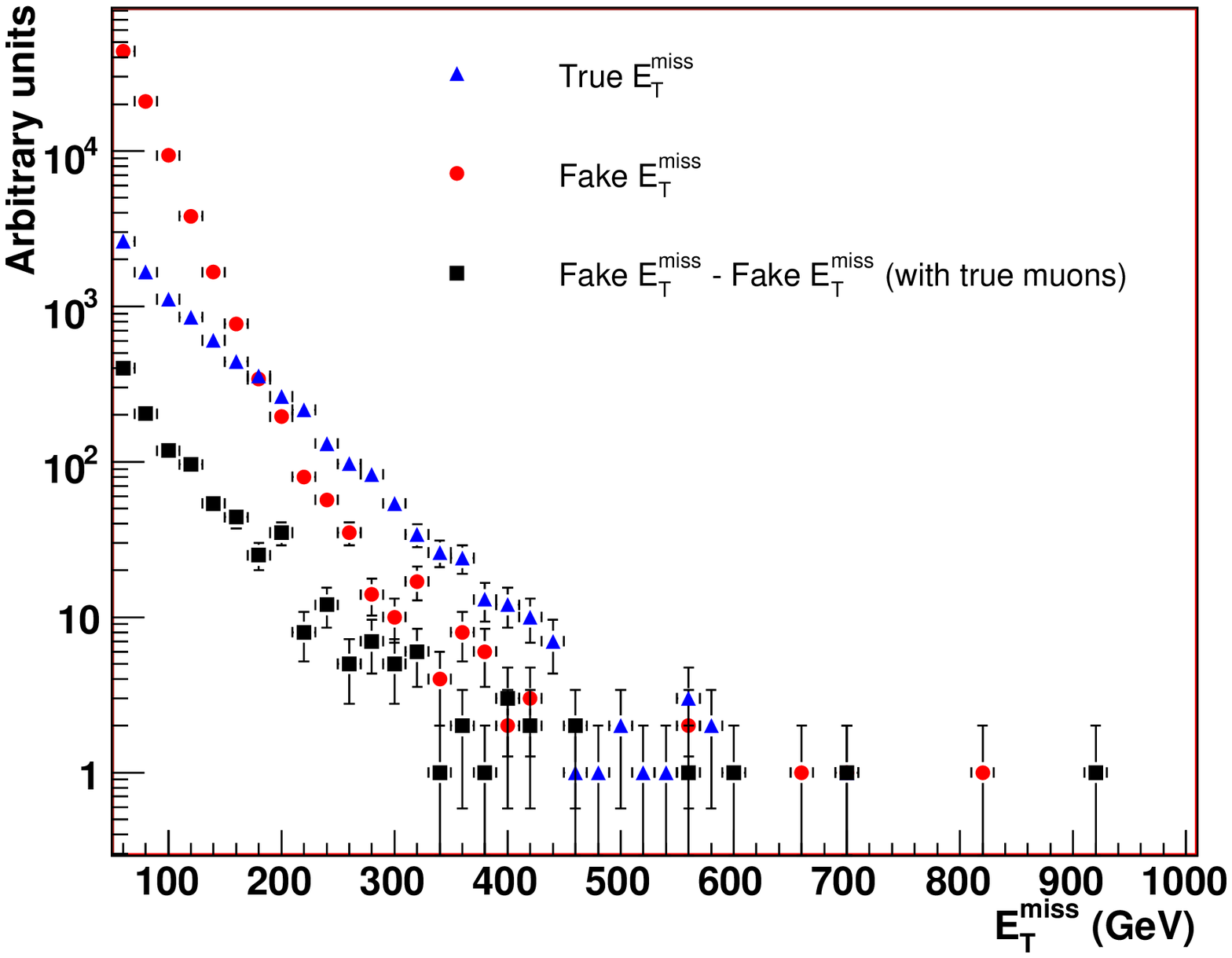}\includegraphics[width=72mm]{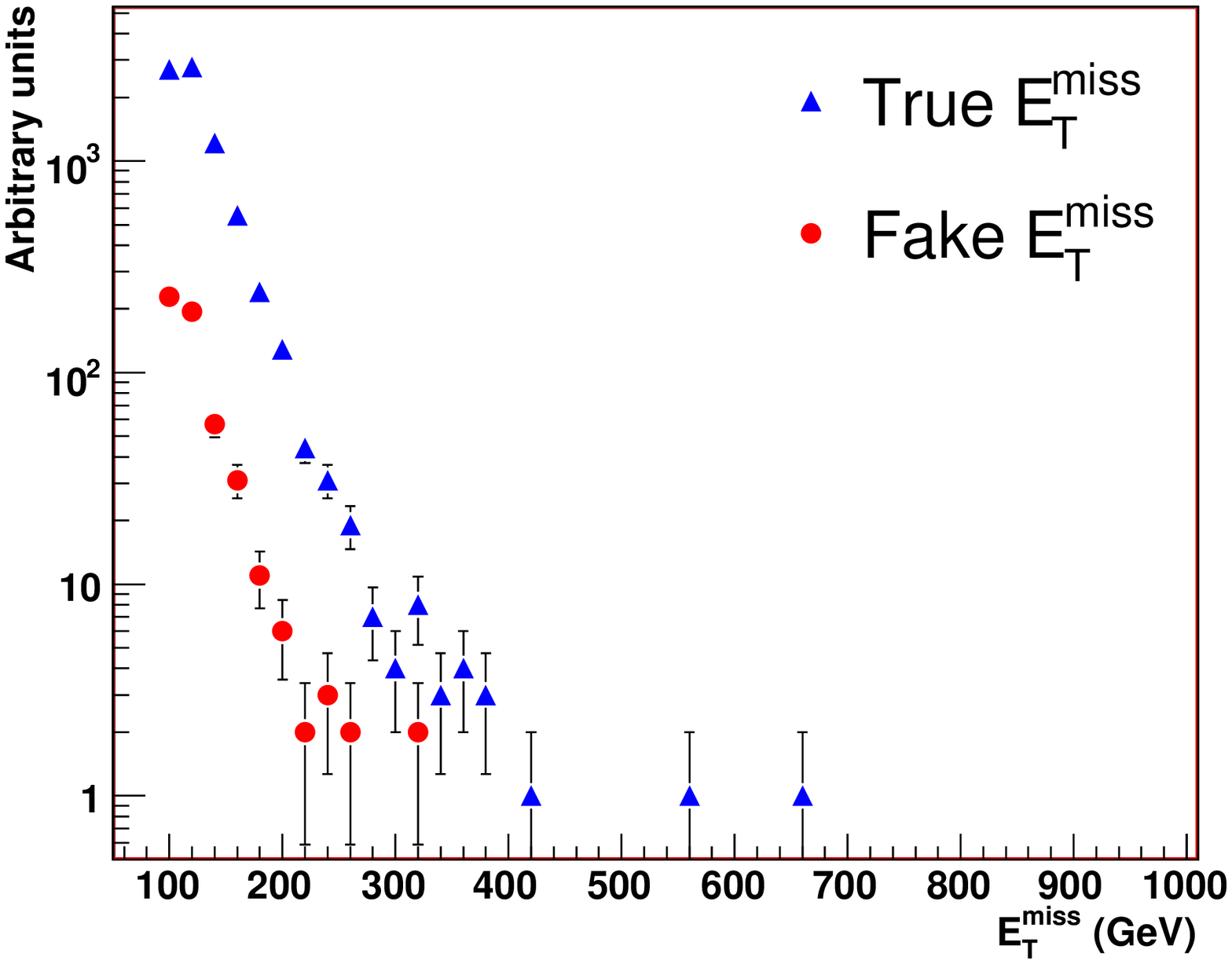}
\caption{Left: Distribution of \MET~variable for generator information 
(triangle) and for fake \MET~(round). The fake \MET~from muon escaping 
the acceptance of ATLAS detector is compared to the overall fake 
\MET~(black point). Right: Distribution of \MET~variable for generator 
information (triangle) and for fake \MET~once the \MET~is corrected from 
muons escaping the acceptance of ATLAS detector (round).} 
\label{fig:MissingEt}
\end{figure*}

Even thought the missing E$_T$ (\MET) variable can be calculated as soon
as the detectors are close and ready, it will take some time before to
fully understand it. In the meantime, \MET~is one of the variable the
most sensitive to new physics. Sources of fake \MET~are mainly beam
gas interactions, dead/hot/noisy cells/area in the calorimeter
systems, non linearity/non-compensated detectors, finite energy
resolution or from muons escaping detection as shown in Fig.~\ref{fig:MissingEt}~\cite{ATLASMET}.

ATLAS and CMS collaboration develop techniques to handle the bias, for
example, coming from jet energy scale
correction~\cite{CMSMet}. Techniques are also developed in order to
estimate the \MET~coming from physics events as $Z^0\rightarrow
\nu\nu$ from ``visible processes'' as $Z^0\rightarrow ll$.

\section{PHYSICS WITH LEPTONS}

\begin{figure*}[t]
\centering
\includegraphics[width=135mm]{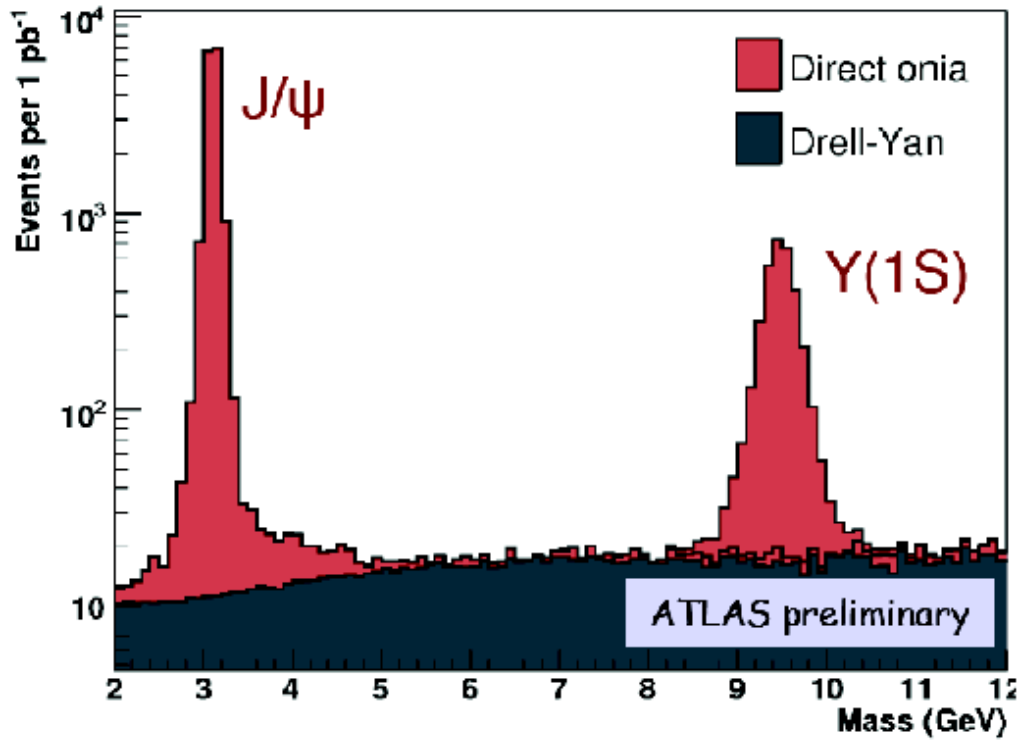}
\caption{Di lepton invariant mass of the expected first inverse 
picobarns of data collected by ATLAS detector. J$/\Psi$ peak as 
well as $\Upsilon$ are clearly visible.} 
\label{fig:Onia}
\end{figure*}

Jets and \MET~are mainly calorimetric object. The combination of the
tracker with other detectors will bring the leptons into the game. But
the leptons will mainly rely on the quality of the alignment of the
tracker and also on the knowledge of magnetic fields. These quantities
can be quickly estimated by looking at resonances as $J/\Psi$ and
$\Upsilon$ within 1 pb$^{-1}$ of integrated luminosity as shown in
Fig.~\ref{fig:Onia}.

The LHC will be a $W^{\pm}$/$Z^0$ factories so with a few amount of
data collected, cross section production can be established for a
energy in the center of mass of 14~TeV. The main studies that will be
done at the beginning with these events will be the commissioning
ones. Indeed, $Z^0$ resonances are really important to tune the
EM-scale of the calorimeter or to control electromagnetic calorimeter
calibration in case of a decay in the electronic channel, to
improve/validate the alignment of tracker and muons chambers in case
of a decay muonically. The resonances will be also used to determine
the efficiencies of the leptons by using the Tag-and-Probe
method~\cite{CMST&P}. The Tag-and-Probe method rely on well known
resonances. One of the leg of the resonances is asked to have a
perfectly well identify lepton. The invariant mass of the two leptons
should be within the resonance mass window. In that case the second
lepton has a high probability to be a true isolated lepton and it is
affected only by the bias from the kinematics of the resonance
production. This second lepton sample can then be used to study lepton
identification criteria.

In addition to cross section measurement, it will be possible to
improve the constraints on PDF as at LHC, the $W^{\pm}$ allow to access
low-x range. With a few events, a measurement with a precision lower
than 5\%, a gain as large as 40\% on systematics can be expected on
low-x gluon shape~\cite{AtlasPDF}.

\begin{figure*}[t]
\centering
\includegraphics[width=135mm]{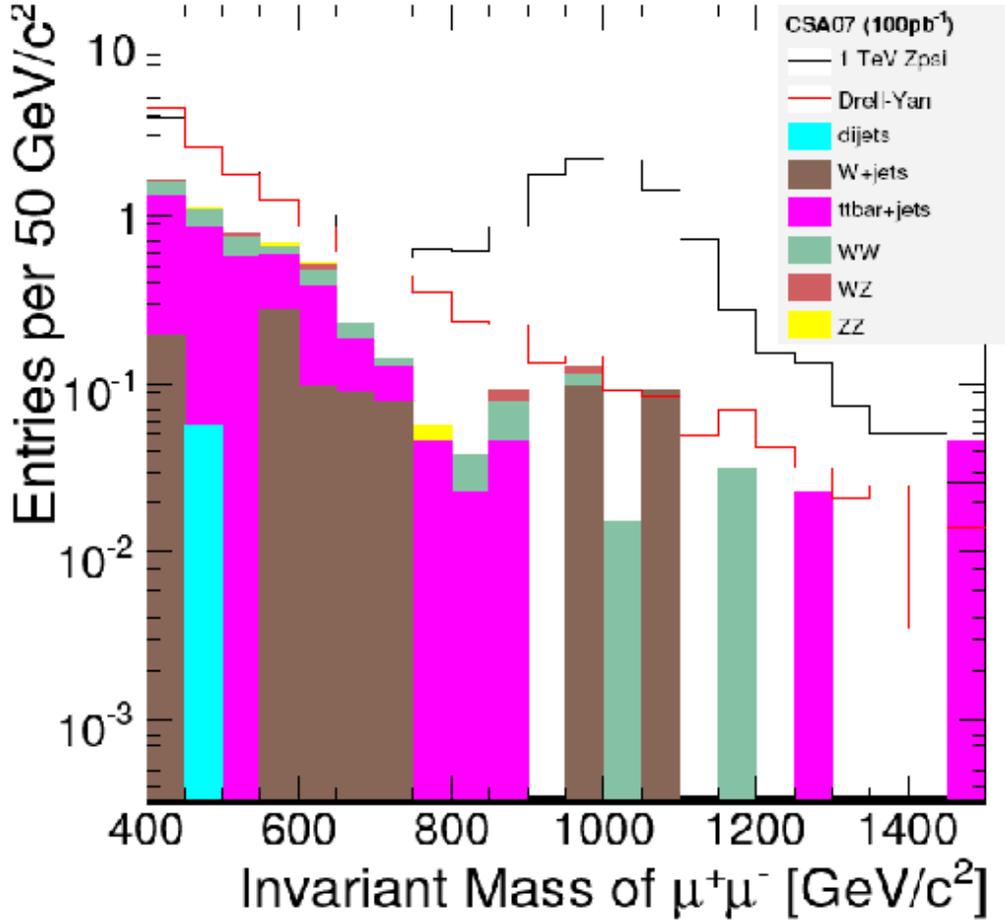}
\caption{Invariant mass of two opposite charge muons within CMS detector 
for an expected luminosity of 100~pb$^{-1}$. A clear signal of $Z_{\psi}$ 
will be seen on top of the different background.} 
\label{fig:MuMu}
\end{figure*}

At this stage, Standard Model physics using leptons as main final
state will be established. With a few inverse picobarn of data, main
properties of such events at a center of mass of 14~TeV will be
studied.  Some early discoveries can happen in these topologies by
simply looking at the tail of the invariant mass distribution. For
example, CMS analysis with 100pb$^{-1}$ of data can see an excess of
events in the tail of the invariant mass of two opposite signed muons
as shown in Fig.~\ref{fig:MuMu}. For such analysis an optimal
detector is not mandatory but the interpretation of such signal as a
possible $Z'$ candidate or as a graviton will require more studies.

\section{TOP PHYSICS}

The $t\bar{t}$ cross section production is a factor 100 greater than
the one at Tevatron. The statistics of such events will be without
comparison and so it will be possible to used these very complex
events in order to validate/perform calibration. From the
semi-leptonic decay of a $t\bar{t}$ event, it will be possible to
constraint the two non b-jet to the invariant mass of the $W$ and in
that case to improve the jet energy scale. The presence of b-jet in
dileptonic decay of such events will give us a high quantity of
events with a high purity of b-jets in order to study b-tagging
efficiency. The
\MET~can also be controlled by constraining the semi-leptonic decay to
the $W$ mass. All these studies can be performed once only top events
are seen in ATLAS and CMS detector. CMS analysis~\cite{CMSTop} shown
that with 10pb$^{-1}$ of collected data, the first measurement of the
$t\bar{t}$ cross section production can be done. The analysis have
been performed for three channels of dileptonic decay of $t\bar{t}$
events (di-electron, dimuon and one electron plus one muon) and in the
case of $W\rightarrow mu \nu$ for semileptonic decay.

\begin{figure*}[t]
\centering
\includegraphics[width=72mm]{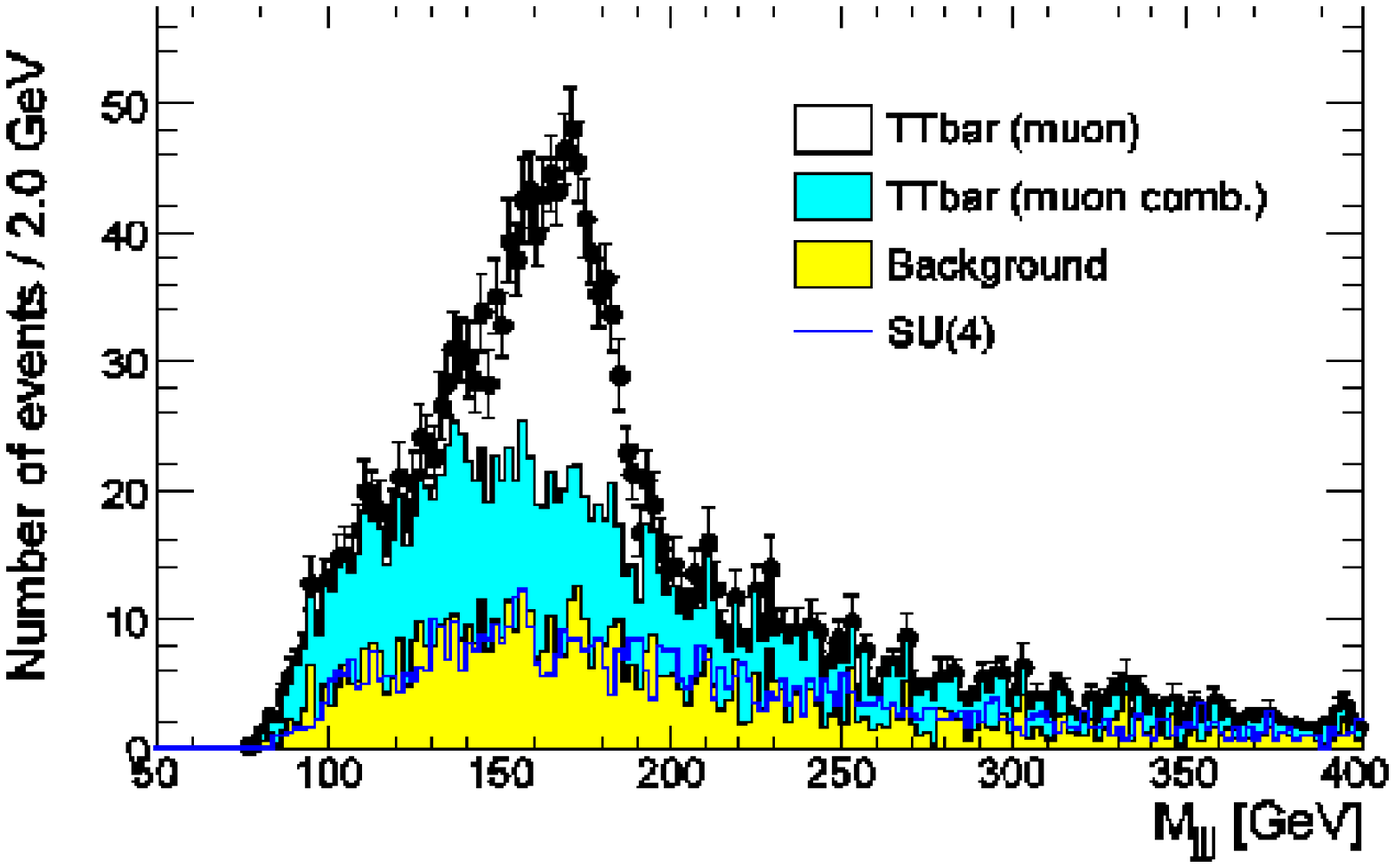}\includegraphics[width=72mm]{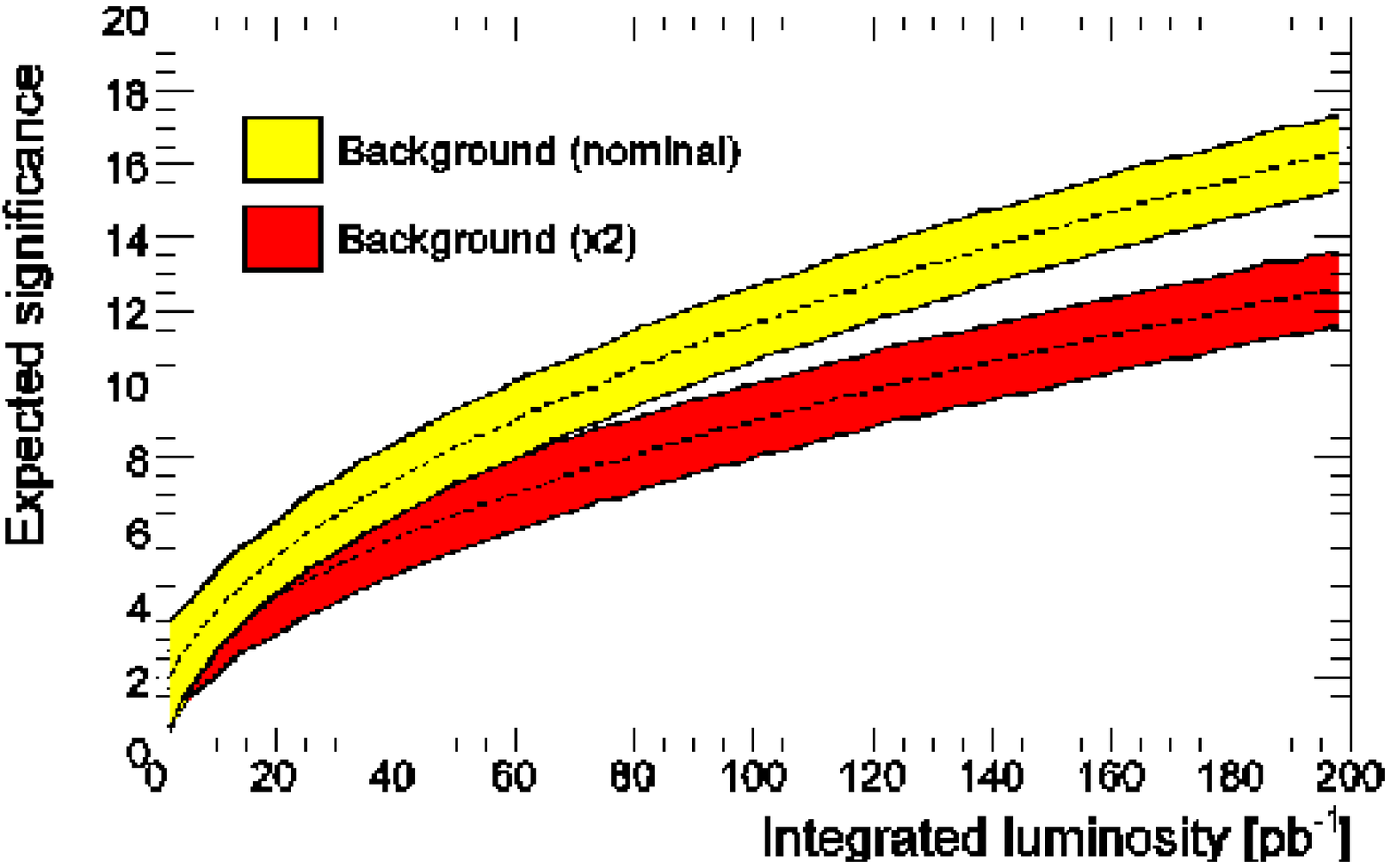}
\caption{Left: Invariant mass of 3 jets in semileptonic $t\bar{t}$ events 
for 100pb$^{-1}$ of expected luminosity recorded by ATLAS detector. 
SU4 signal will double the level of background. Right: Significance 
as a function of integrated luminosity with and without a SU4 signal 
which will double the background.} 
\label{fig:Su4}
\end{figure*}

Physics beyond Standard Model can also play an important role in the
studies of $t\bar{t}$ signal at LHC. ATLAS analysis studied a squark
signal within mSugra framework which will appear as doubling the
background for $t\bar{t}$ semi-leptonic analysis. The invariant mass
of 3 jets as well as significance as function of integrated luminosity
are presented in Fig.~\ref{fig:Su4}.

Once $t\bar{t}$ production is established at a center of mass of 14~TeV
studies of the invariant mass of the $t\bar{t}$ system can be
scrutinized. An excess in such distribution can give a hint towards
beyond Standard Model processes.

Contrary to $t\bar{t}$ system, the production of single top will
require a integrated luminosity of around 1fb$^{-1}$ to be
established. Nevertheless, this processes will allow us to test quite
some theories. The cross section production can be enlarged in case of
$b', t'$ production if $M_{b'} > M_{t'}$, $W'$ production, flavor
changing neutral current as well as Susy correction etc.

The top physics is a really challenging one due to the complex final
state but this also conclude the commissioning phase of the detectors
and allow us to push forward discoveries.

\section{HIGGS BOSON SEARCH}
\begin{figure*}[t]
\centering
\includegraphics[width=135mm]{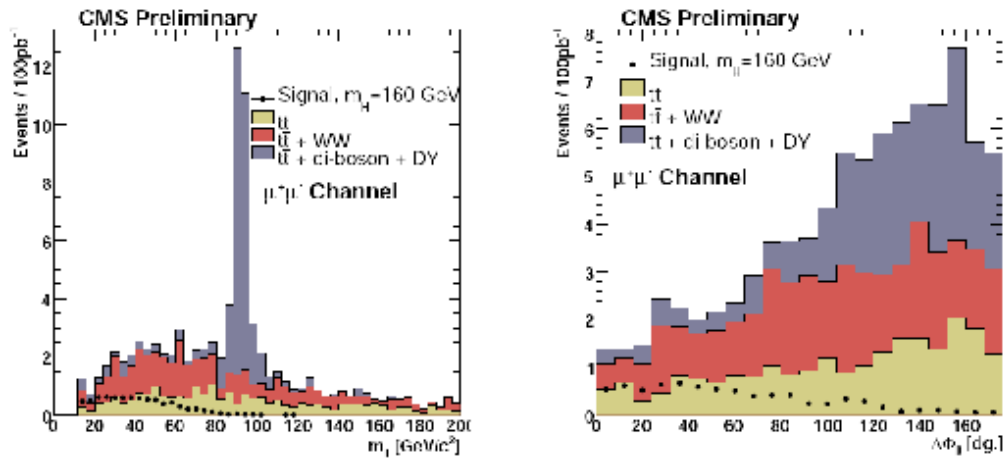}
\caption{Left: Invariant mass of the dilepton system reconstructed in the CMS detector, right: azimuthal 
angular separation between the two leptons for the $\mu^{\pm}\mu^{\mp}$ 
channel after the selection and for a Standard Model Higgs boson mass
 hypothesis of $m_H = 160$ GeV.}
\label{fig:Higgs}
\end{figure*}

The last remaining piece of the Standard Model which is still missing
is the search for a Higgs signal. Depending on the mass of the Higgs
boson, the Higgs boson can be discovered in a really clean and
controlled multi lepton final state with a few hundreds of picobarns of
data if the Higgs boson mass is larger than 130 GeV. In that case the
Higgs boson decay will be essentially via a pair of $W$ boson. These
bosons will be studied in their leptonic decay and mainly electron and
muon. Due to the large \MET~expected from the $W$ boson decay, the
mass of Higgs boson cannot be reconstructed. The Higgs boson is a spin
0 particle so the two leptons coming from $W$ decay will tend to be
collinear. This property gives a handle to discriminate Higgs boson
decay from $WW$ Standard Model production. Fig.~\ref{fig:Higgs}
presents for CMS analysis, the invariant mass of the dilepton system
and the azimuthal angular separation between the two leptons for the
$\mu^{\pm}\mu^{\mp}$ channel after the selection applied and for a
Standard Model Higgs boson mass hypothesis of $m_H =
160$~GeV~\cite{CMSHiggs}

The Higgs boson discovery or its exclusion will play a crucial role
for the seacrhes beyond Standard Model physics if by that time nothing
would have been already observed.

\section{CONCLUSION}

\begin{figure*}[t]
\centering
\includegraphics[width=135mm]{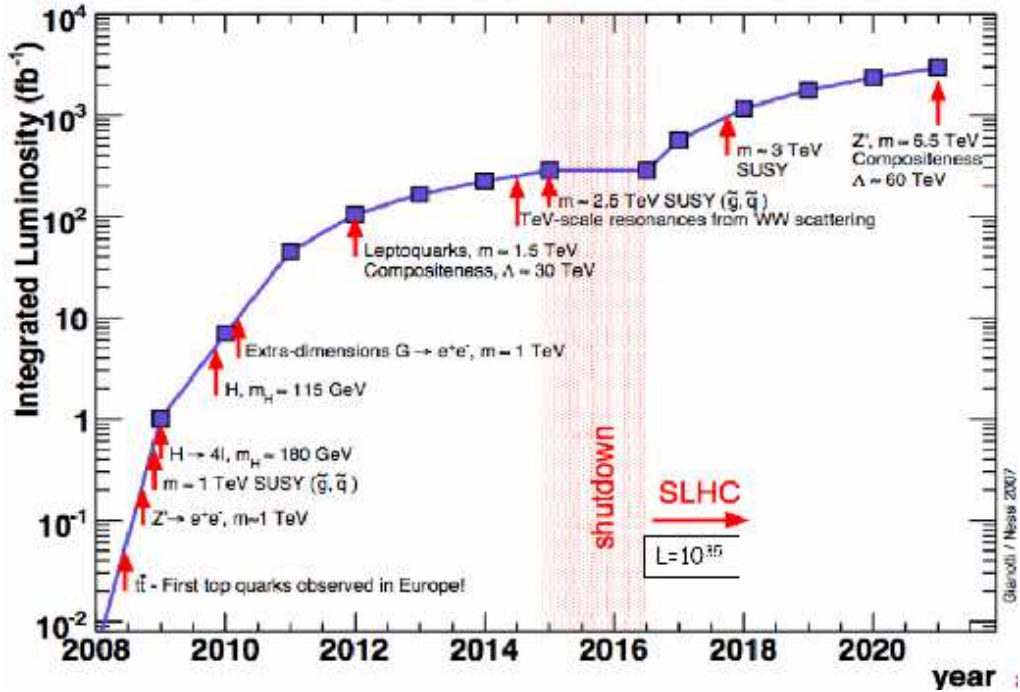}
\caption{Integrated Luminosity as a function of time with expected point 
where discovery of process beyond Standard Model can be performed. 
LHC will allow us to achieve large searches.} 
\label{fig:Conclu}
\end{figure*}

Once the LHC will start and once ATLAS and CMS experiment will have
commissioned their detectors using Standard Model processes, the vast
area of searches for processes beyond Standard Model will really
begin. Some processes can already appear within the commissioning
phase or at least indicate that beyond Standard Model physics is at
the corner. Fig.~\ref{fig:Conclu} presents as a function of time and
integrated luminosity the discovery potential owing to LHC machine and
detectors.

\begin{acknowledgments}
I would like to thank the conference organisers for the invitation to give this
talk. Thanks, also, to ATLAS and CMS colleagues for their input material.

\end{acknowledgments}


\begin{thebibliography}{9}   

\bibitem{Atlas}
ATLAS Collaboration, ``ATLAS detector and physics performance'', Technical Design Report,
CERN-LHCC-99-15/CERN-LHCC-99-14 (1999)

\bibitem{CMS}
CMS Collaboration, ``The CMS experiment at the CERN LHC'', Submitted to Journal of
Instrumentation (2008)

\bibitem{CMSContactInteraction}
CMS Collaboration, ``Searches for New Physics using high ET dijet events'', CMS PAS SBM-07-001 (2007)
 
\bibitem{CMSMet}
CMS Collaboration, ``Missing ET performance in CMS'', CMS PAS JME-07-001 (2007)

\bibitem{ATLASMET}
ATLAS Collaboration, ``The ATLAS Experiment at the CERN Large Hadron Collider'', 2008 JINST 3 S08003

\bibitem{CMST&P}
CMS Collaboration, ``Measuring Electron Efficiencies at CMS with Early Data'', CMS PAS EGM-07-001 (2007)

\bibitem{AtlasPDF}
A.~Tricoli, A.~Cooper-Sarkar and C.~Gwenlan, ``Uncertainties on W and Z production at the LHC'', hep-ex/0509002

\bibitem{CMSTop}
CMS Collaboration, ``Observability of Top Quark Pair Production in the Semileptonic Muon Channel with the first 10 pb$^{-1}$ of CMS Data'', CMS PAS TOP-08-005 (2008)


\bibitem{CMSHiggs}
CMS Collaboration, ``Search for the Higgs boson in the $WW^{(*)}$ decay channel with the CMS experiment'', CMS PAS Hig-07-001 (2007)

\end{thebibliography}
\end{document}